\documentclass[aps,superscriptaddress,amsmath,amssymb]{revtex4}

\usepackage[applemac]{inputenc}
\usepackage[T1]{fontenc}
\usepackage{amsfonts, amssymb, amsmath, dsfont}
\usepackage{graphicx}
\usepackage{float}
\usepackage{ulem}

\begin{document}

\title{Comment on arXiv:1705.01165 by V. N. Premakumar, M. G. Vavilov and R. Joynt}

\author{R\'emi Carminati}
\email{remi.carminati@espci.fr}
\affiliation{ESPCI Paris, PSL Research University, CNRS, Institut Langevin, 1 rue Jussieu, F-75005, Paris, France.}
\author{Boris Shapiro}
\email{boris@physics.technion.ac.il} \affiliation{Department of
Physics, Technion, 32000 Haifa, Israel.}
\begin{abstract}

\end{abstract}

\maketitle

The authors compute various correlations in a fluctuating thermal
electromagnetic field, close to the surface of a material body. We
would like to point out that many of the results derived by the
authors have been known for many years and, on top of being
published in original articles, are even often documented in
books~\cite{Levin_book,Rytov_book} and
reviews~\cite{Joulain_2005}. Thus, instead of lengthy derivations,
the authors could have copied the relevant equations from the
early sources, with an appropriate reference. For example,
authors' Eq. (42) appears in
\cite{Levin_book,Rytov_book,Joulain_2005}, while their results for
"extended qubits" (the unnumbered equations on p. 14) can be found
in Ref.~\cite{Henkel_2000}. Geometries other than half-space
(cylinders, spheres) have also been considered in the
literature~\cite{Levin_book,Rytov_book}, and the same goes for the
"magnetic noise" problem~\cite{Agarwal_1975,Joulain_2003}.

Some important developments in the topic of near-field electromagnetic fluctuations 
are also either addressed superficially, or simply ignored. For example, in a short paragraph (section 2.4), 
the authors comment on the failure of the local material response at very short distance from surfaces, 
and on the fact that nonlocality needs to be included to correct for unphysical divergences. 
They write that ``... to date only the problems of a conducting half-space 
 and conducting films have been treated using the nonlocal formalism'', and cite their own recent works. 
 Actually, the question of nonlocality and the treatment of divergences has been discussed in textbooks~\cite{Levin_book,Rytov_book},
 and precise asymptotic expressions of field correlation functions using nonlocal response have been published~\cite{Henkel_2006}. 
 Furthermore, the topic of thermal electromagnetic fluctuations is now 
 at a stage where theoretical predictions can be compared to experiments. Indeed, specific near-field microscopy 
 techniques have been developed to directly measure near-field thermal fluctuations~\cite{DeWilde_2006}. Even in a theoretical
 paper it could be useful to briefly mention the connection with state-of-the-art experimental techniques. For example, near-field spectra 
 of the energy density of the fluctuating electric field [corresponding to the quantity in Eq.~(43)] have been measured~\cite{DeWilde_2013}.
 
In addition, in its introductory part the paper contains misleading and inaccurate statements. The authors write that the power radiated by a hot body depends only on the surface area and temperature of the body. It is well known, however, that the surface emissivity is also an important factor. Radiation from hot bodies into a cold environment is a well studied problem, and the authors' discussion of some "paradox" and its resolution seems a bit misterious.

In conclusion, the fluctuational electrodynamics, which the authors use in their study of qubit decoherence, is a well established field and there is no need to derive from scratch the results which can be easily found in the literature.


\begin{thebibliography}{20}

\bibitem{Levin_book}
M.L. Levin and S.M. Rytov, {\it Theory of Equilibrium Thermal
Fluctuations in Electrodynamics} (Nauka, Moscow 1967) (in Russian).

\bibitem{Rytov_book}
S.M. Rytov, Yu.A. Kravtsov and V.I. Tatarskii, {\it Principles of
Statistical Radiophysics}, Vol. 3, ch. 3 (Springer, Berlin, 1989).

\bibitem{Joulain_2005}
K. Joulain, J.-P. Mulet, F. Marquier, R. Carminati and J.-J.
Greffet, Surf. Sci. Rep. \textbf{57}, 59 (2005).

\bibitem{Henkel_2000}
C. Henkel, K. Joulain, R. Carminati and J.-J. Greffet, Opt. Commun. {\bf 186}, 57 (2000).

\bibitem{Agarwal_1975}
G.S. Agarwal, Phys. Rev. A {\bf 11}, 230 (1975).

\bibitem{Joulain_2003}
K. Joulain, R. Carminati, J.P. Mulet and J.J. Greffet, Phys. Rev. B {\bf 68}, 245405 (2003).

\bibitem{Henkel_2006}
C. Henkel and K. Joulain, Appl. Phys. B {\bf 84}, 61 (2006).

\bibitem{DeWilde_2006}
Y. De Wilde, F. Formanek, R. Carminati, B. Gralak, P.A. Lemoine, K. Joulain, J.P. Mulet, Y. Chen and J.J. Greffet,
Nature {\bf 444}, 740 (2006).

\bibitem{DeWilde_2013}
A. Babuty, K. Joulain, P.-O. Chapuis, J.-J. Greffet, Y. De Wilde, Phys. Rev. Lett. {\bf 110}, 146103 (2013).


\end{thebibliography}
\end{document}